\documentclass[a4paper,11pt]{article}
\pdfoutput=1 

\usepackage{jinstpub} 

\usepackage{caption}
\usepackage{subcaption}
\usepackage{verbatim}
\usepackage[symbol]{footmisc}
\usepackage[normalem]{ulem}
\newcommand\redout{\bgroup\markoverwith
{\textcolor{red}{\rule[0.5ex]{2pt}{0.8pt}}}\ULon}

\title{\boldmath The JSNS$^{2}$ data acquisition system}

\author[1,]{J. S. Park\footnote[1]{Corresponding author. e-mail : jspark@post.kek.jp}}
\author[4]{S.~Ajimura} 
\author[9]{M.~Botran}
\author[5]{M.~K.~Cheoun}
\author[6]{J.~H.~Choi}
\author[7]{T.~Dodo}
\author[1]{H.~Furuta}
\author[3]{P.~Gwak}
\author[8]{M.~Harada} 
\author[8]{S.~Hasegawa} 
\author[7, 8]{Y.~Hino}
\author[4]{T.~Hiraiwa} 
\author[11]{H.~I.~Jang}
\author[2]{J.~S.~Jang}
\author[3]{M.~Jang}
\author[19]{H.~Jeon}
\author[19]{S.~Jeon}
\author[3]{K.~K.~Joo}
\author[9]{J.~R.~Jordan} 
\author[19]{D.~E.~Jung}
\author[12]{S.~K.~Kang}
\author[8]{Y.~Kasugai} 
\author[10]{T.~Kawasaki} 
\author[13]{E.~J.~Kim}
\author[3]{J.~Y.~Kim}
\author[19]{S.~B.~Kim}
\author[14]{S.~Y.~Kim}
\author[15]{W.~Kim}
\author[10]{T.~Konno}
\author[19]{E.~Kwon}
\author[19]{D.~H.~Lee}
\author[3]{I.~T.~Lim}
\author[1]{T.~Maruyama}
\author[9]{E.~Marzec} 
\author[8]{S.~Meigo} 
\author[1]{S.~Monjushiro}
\author[3]{D.~H.~Moon}
\author[4]{T.~Nakano} 
\author[16]{M.~Niiyama}
\author[1]{K.~Nishikawa\footnote[2]{deceased}} 
\author[4]{M.~Nomachi} 
\author[6]{M.~Y.~Pac}
\author[17]{S.~J.~M.~Peeters}
\author[18]{H.~Ray}
\author[19]{C.~Rott} 
\author[8]{K.~Sakai} 
\author[8]{S.~Sakamoto}
\author[14]{H.~Seo} 
\author[3]{J.~H.~Seo}
\author[4]{T.~Shima} 
\author[3]{C.~D.~Shin}
\author[9]{J.~Spitz} 
\author[20]{I.~Stancu}
\author[7]{F.~Suekane}
\author[4]{Y.~Sugaya}
\author[8]{K.~Suzuya}
\author[1]{M.~Taira}
\author[7]{R.~Ujiie}
\author[21]{M.~Yeh}
\author[19]{I.~Yu} 
\author[3]{A.~Zohaib}

\affiliation[1]{\it{\footnotesize{High Energy Accelerator Research Organization (KEK), Tsukuba, Ibaraki, JAPAN}}}
\affiliation[2]{\it{\footnotesize {GIST College, Gwangju Institute of Science and Technology, Gwangju, 61005, KOREA}}}
\affiliation[3]{\it{\footnotesize {Department of Physics, Chonnam National University, Gwangju, 61186, KOREA}}}
\affiliation[4]{\it{\it{\footnotesize {Research Center for Nuclear Physics, Osaka University, Osaka, JAPAN}}}}
\affiliation[5]{\it{\footnotesize {Department of Physics, Soongsil University, Seoul 06978, KOREA}}}
\affiliation[6]{\it{\footnotesize{Department of Radiology, Dongshin University, Chonnam 58245, KOREA}}}
\affiliation[7]{\it{\footnotesize {Research Center for Neutrino Science, Tohoku University, Sendai, Miyagi, JAPAN}}}
\affiliation[8]{\it{\footnotesize {J-PARC Center, JAEA, Tokai, Ibaraki JAPAN}}}
\affiliation[9]{\it{\footnotesize {University of Michigan, Ann Arbor, MI, 48109, USA}}}
\affiliation[10]{\it{\footnotesize{Department of Physics, Kitasato University, Sagamihara 252-0373, Kanagawa, JAPAN}}}
\affiliation[11]{\it{\footnotesize {Department of Fire Safety, Seoyeong University, Gwangju 61268, KOREA}}}
\affiliation[12]{\it{\footnotesize{School of Liberal Arts, Seoul National University of Science and Technology, Seoul, 139-743, KOREA}}}
\affiliation[13]{\it{\footnotesize {Division of Science Education, Physics major, Chonbuk National University, Jeonju, 54896, KOREA}}}
\affiliation[14]{\it{\footnotesize{Department of Physics and Astronomy, Seoul National University, Seoul 08826, KOREA}}}
\affiliation[15]{\it{\footnotesize{Department of Physics, Kyungpook National University, Daegu 41566, KOREA}}}
\affiliation[16]{\it{\footnotesize{Department of Physics, Kyoto Sangyo University, Kyoto, JAPAN}}}
\affiliation[17]{\it{\footnotesize{Department of Physics and Astronomy, University of Sussex, Brighton,UK}}}
\affiliation[18]{\it{\footnotesize {University of Florida, Gainesville, FL, 32611, USA}}}
\affiliation[19]{\it{\footnotesize{Department of Physics, Sungkyunkwan University, Suwon 16419, KOREA}}}
\affiliation[20]{\it{\footnotesize {University of Alabama, Tuscaloosa, AL, 35487, USA}}}
\affiliation[21]{\it{\footnotesize {Brookhaven National Laboratory, Upton, NY, 11973-5000, USA}}}  

\abstract{The JSNS$^{2}$ (J-PARC Sterile Neutrino Search at J-PARC Spallation Neutron Source) experiment aims to search for neutrino oscillations over a 24 m short baseline at J-PARC. The JSNS$^{2}$ inner detector is filled with 17 tons of gadolinium(Gd)-loaded liquid scintillator (LS) with an additional 31 tons of unloaded LS in the intermediate $\gamma$-catcher and an optically separated outer veto volumes. A total of 120 10-inch photomultiplier tubes observe the scintillating optical photons and each analog waveform is stored with the flash analog-to-digital converters. We present details of the data acquisition, processing, and data quality monitoring system. We also present two different trigger logics which are developed for the beam and self-trigger. }


\begin{document}
\maketitle
\flushbottom

\section{Introduction}
The JSNS$^{2}$ (J-PARC Sterile Neutrino Search at J-PARC Spallation Neutron Source) experiment searches for neutrino oscillations at a 24\,m baseline with $\Delta$m$^{2}$ near 1\,eV$^{2}$ at the J-PARC Material and Life Science Experimental Facility (MLF)~\cite{cite:JSNS2_proposal}. A 3\,GeV proton beam is directed onto a mercury target to produce muons. The muons are stopped by the steel shielding near the target where they decay producing decay-at-rest neutrinos. JSNS$^{2}$ measures the $\overline{\nu}_{\mu}$ $\rightarrow$ $\overline{\nu}_{e}$ oscillations, which can be detected via inverse beta decay (IBD), $\overline{\nu}_{e} + p \rightarrow e^{+} + n$. The detector consists of an inner target volume, an intermediate $\gamma$-catcher region, and an outer veto~\cite{cite:JSNS2_TDR, cite:JSNS2_Veto_detector}. The target region is filled with 17 tons of gadolinium(Gd) loaded liquid scintillator (LS). The optically separated $\gamma$-catcher and outer veto are filled with 31 tons of unloaded LS. A total of 120 10-inch photomultiplier tubes (PMTs) were pre-calibrated~\cite{cite:PMT_precalib} and installed in the detector. Of the 120 PMTs, 96 are directed inward to view the target volume and 24 PMTs view the veto. 
The JSNS$^{2}$ data acquisition system (DAQ) reuses 28 CAEN v1721 flash analog-to-digital converter (FADC)~\cite{cite:CAEN_FADC} and 16 front-end electronics (FEE) boards that were donated to us by the Double Chooz experiment~\cite{cite:FEE_DC}, for cost-effective but still suitable for the JSNS$^{2}$ experiment. 
The DAQ records waveforms from the PMTs, transfers the raw data to the storage area. At the same time, the data quality monitoring (DQM) system does a fast processing of the data. 

The JSNS$^{2}$ DAQ must meet several requirements for data taking. The first requirement is that the DAQ can be triggered synchronously with the MLF beam and that several microseconds of data can be readout following each beam event. Beam-target interactions occur at a fixed rate, uniform 25\,Hz rate, meaning that the DAQ must be able to readout 17\,MB/s of data. The next requirement is for detector calibration. Several sources, including natural Michel electrons from cosmogenic muons and a deployed $^{252}$Cf source, will be used to calibrate the detector response. The DAQ must be able to trigger on events from those sources, and so in addition to beam synchronous triggers there must also be "self-trigger" logic. Another requirement is that enough data is readout for each event so that pulse shape discrimination can be used as a method of particle identification. The fraction of scintillation light that appears in a prompt time window is expected to be larger for gamma or positron event than for a neutron event. So the DAQ must readout enough data after each interaction such that the fraction of prompt and late light can be accurately determined. Reliable particle identification is necessary for removing backgrounds from out IBD dataset. In this paper, we present the details of the DAQ/DQM system including hardware\redout{s} in sec.~\ref{sec:DAQ}, two different trigger algorithms in sec.~\ref{sec:trigger}.

\section{Data acquisition system}
\label{sec:DAQ}
\begin{figure}[h]
\begin{center}
\includegraphics[scale=0.5]{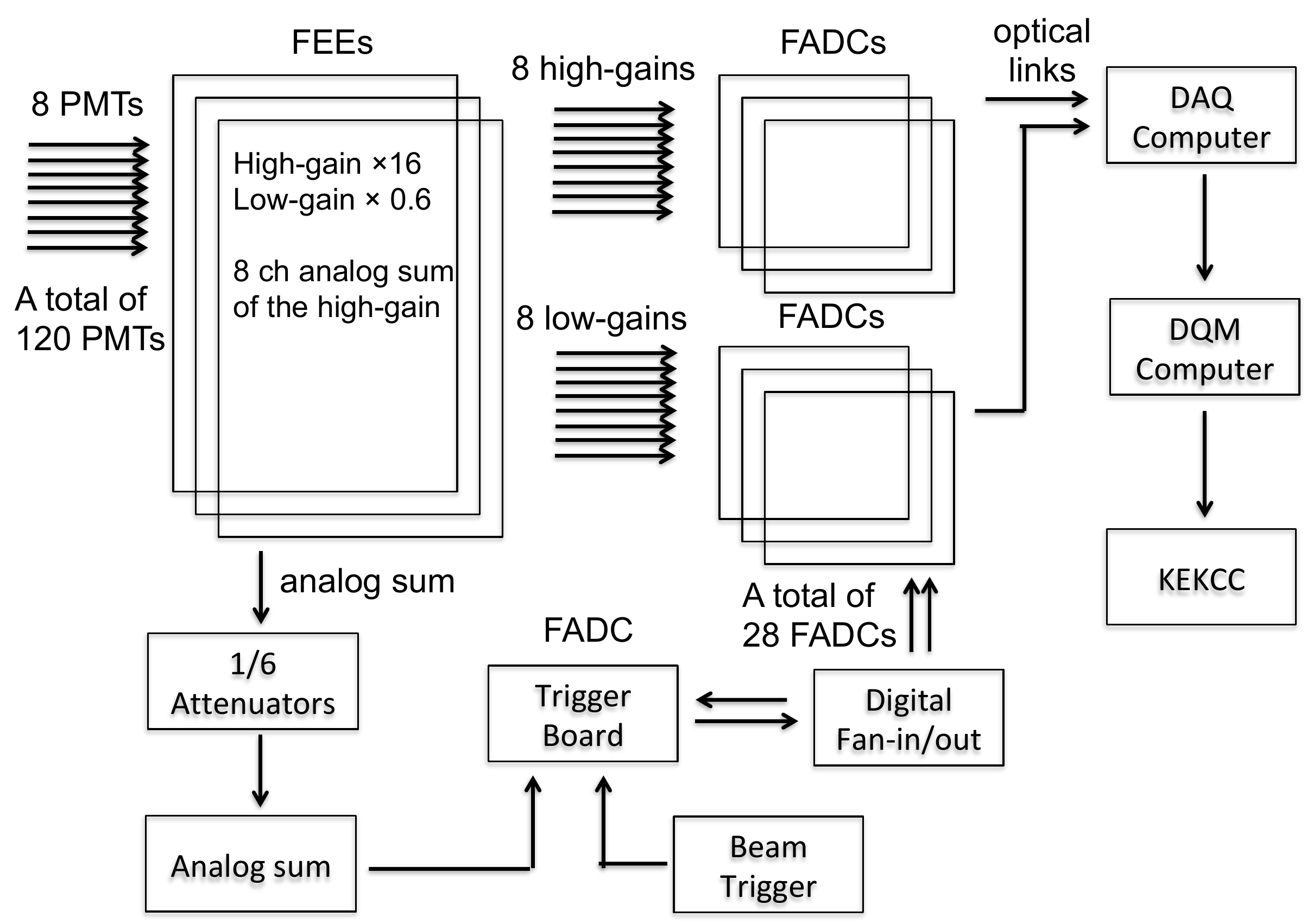}
\end{center}
\caption{\setlength{\baselineskip}{4mm}Data flow for the JSNS$^{2}$ experiment. 8 PMTs signals fed into one FEE, amplified by two different gains, and are digitized by two FADCs. Each FEE generates an analog sum of 8 high-gain signals, fed into an attenuator module which has a factor of 1/6, and summed up with an analog sum module to get a total analog sum of inner 96 PMTs. A FADC trigger board receives both the beam trigger and the analog sum and generates global trigger which fed into a digital Fan-in/out. The copied global triggers are inserted into each FADC including the FADC trigger board.} 
\label{fig:DAQ_scheme}
\end{figure}

Figure~\ref{fig:DAQ_scheme} shows a schematic view of the JSNS$^{2}$ data flow. The DAQ consists of sixteen FEE modules, twenty eight FADC modules, a DAQ computer, and a DQM computer. The FEE splits each PMT signal into two copies, a high-gain copy that is amplified by a factor of 16 and a low-gain copy that is attenuated by a factor of 0.6. Note that JSNS$^{2}$ stores both the high-gain and low-gain of the 96 inner PMTs waveforms, however, stores only high-gain of the 24 outer PMTs. The high-gain and low-gain signals from a single PMT can be recombined in an offline reconstruction or analysis procedure to increase the effective charge resolution; such precise charge resolution is only needed for the inner PMTs. Therefore, a total of 216 channels are recorded by the DAQ system. One FADC board is designated as a trigger board, it generates a global trigger and provides master clock to synchronize the clocks of all the other FADC modules. The trigger FADC generates a global trigger whenever it receives a beam trigger, or if the analog sum of the inner 96 PMTs exceeds a preset threshold.  A DAQ PC reads out data from each FADC module over an optical fibre and transfers the data to a DQM PC via a TCP/IP socket. The DQM PC stores the raw data to its disk, then transfers the raw data to the KEK computing center (KEKCC) for long-term storage and later analysis. Once the transfer is complete and confirmed, the raw data files are erased from the DQM PC to conserve local disk space. In parallel to the data transfer, the DQM PC does a fast data analysis to ensure that each PMT's waveform and charge distribution falls within preset bounds. 

\subsection{Front-end electronics}
The FEE~\cite{cite:FEE_DC} is a NIM module that has an 8-pin Combo-D inputs on the rear panel and two 8-pin Combo-D outputs on the front panel; one connector is for the high-gain output signals and the other connector for the low-gain signals. We modified the amplification factors for the FEE's low-gain channels to achieve a gain of 0.6, a more optimal value for the JSNS$^{2}$ experiment. In addition, as shown in Fig.~\ref{fig:DAQ_scheme}, each FEE provides analog sum of the eight high-gains which is useful to generate triggers.

\begin{figure}[h]
\begin{center}
\includegraphics[scale=0.35]{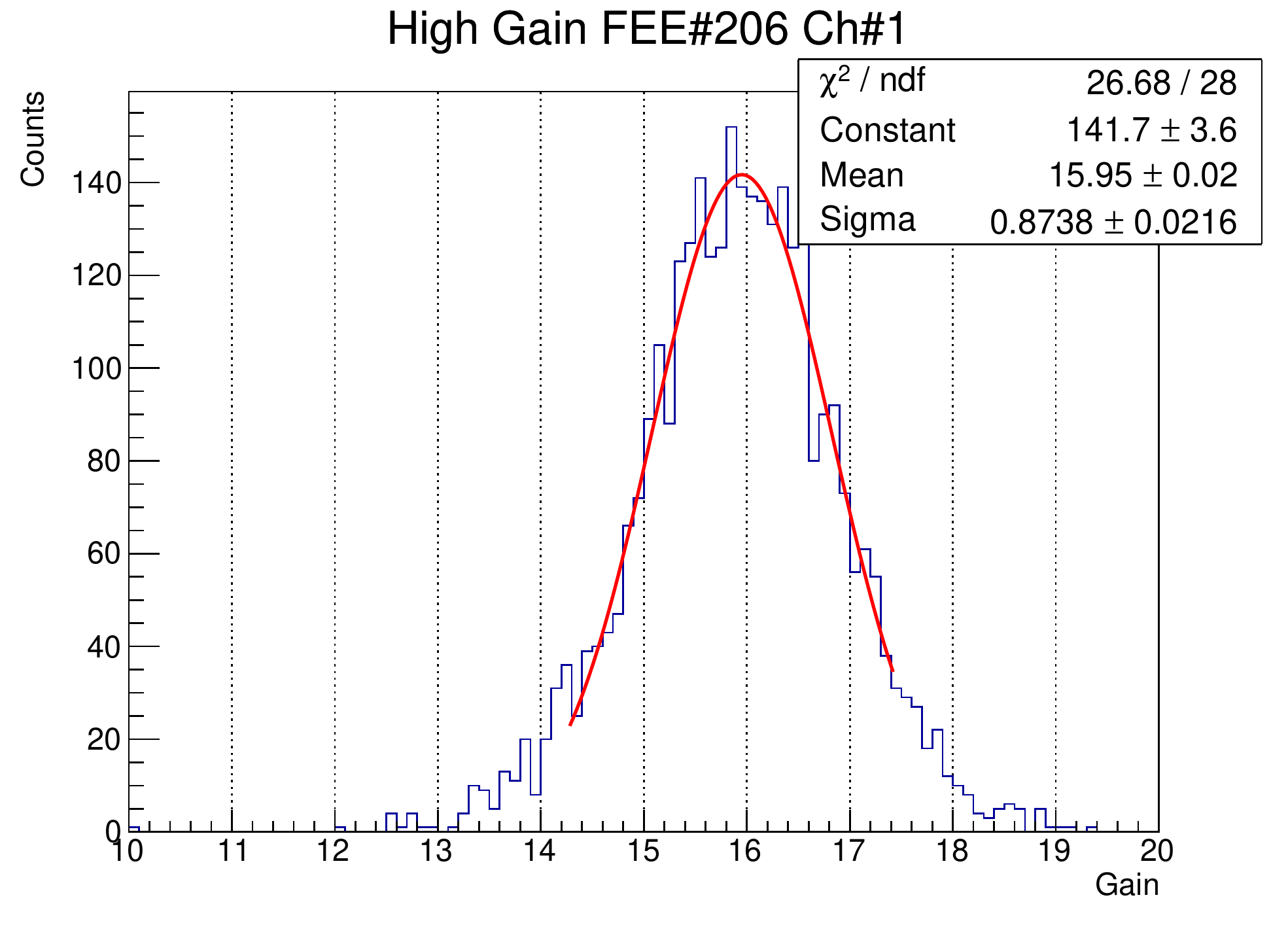}
\includegraphics[scale=0.35]{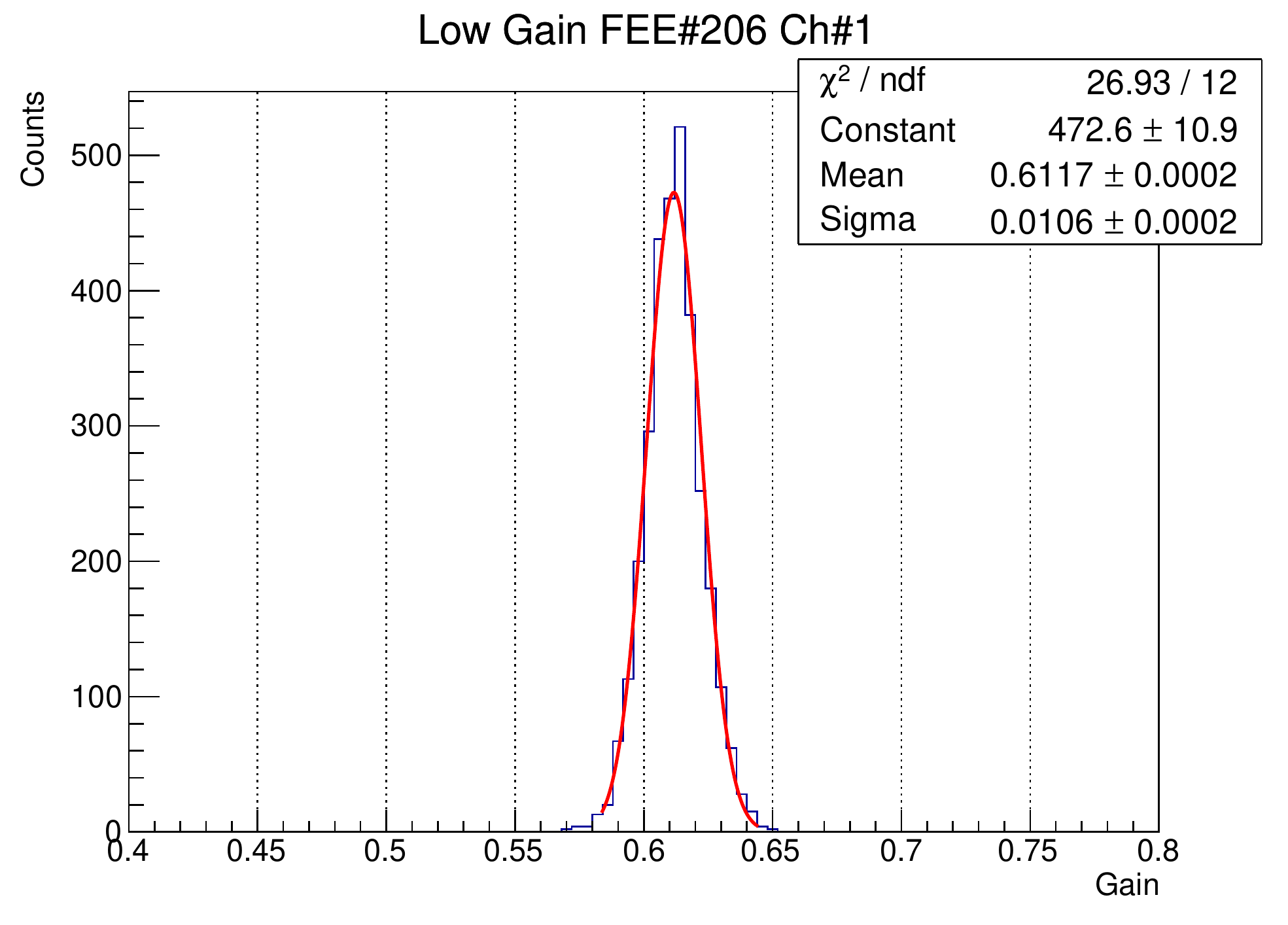}
\end{center}
\caption{\setlength{\baselineskip}{4mm} Example of FEE gain fitting for a single FEE channel, high-gain (left) and low-gain (right).}
\label{fig:FEE_gain_fitting}
\end{figure}

\begin{figure}[h]
\begin{center}
\includegraphics[scale=0.35]{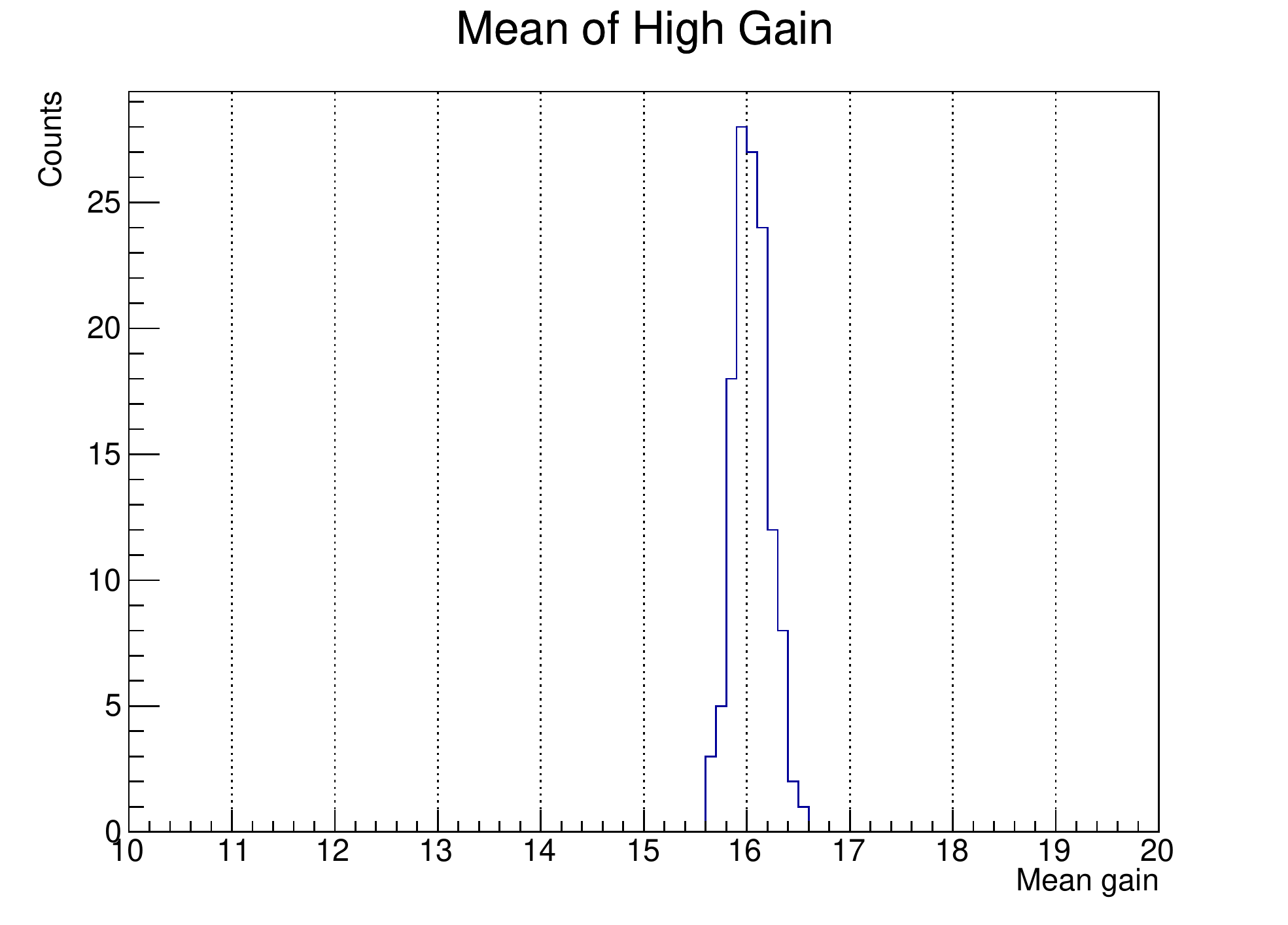}
\includegraphics[scale=0.35]{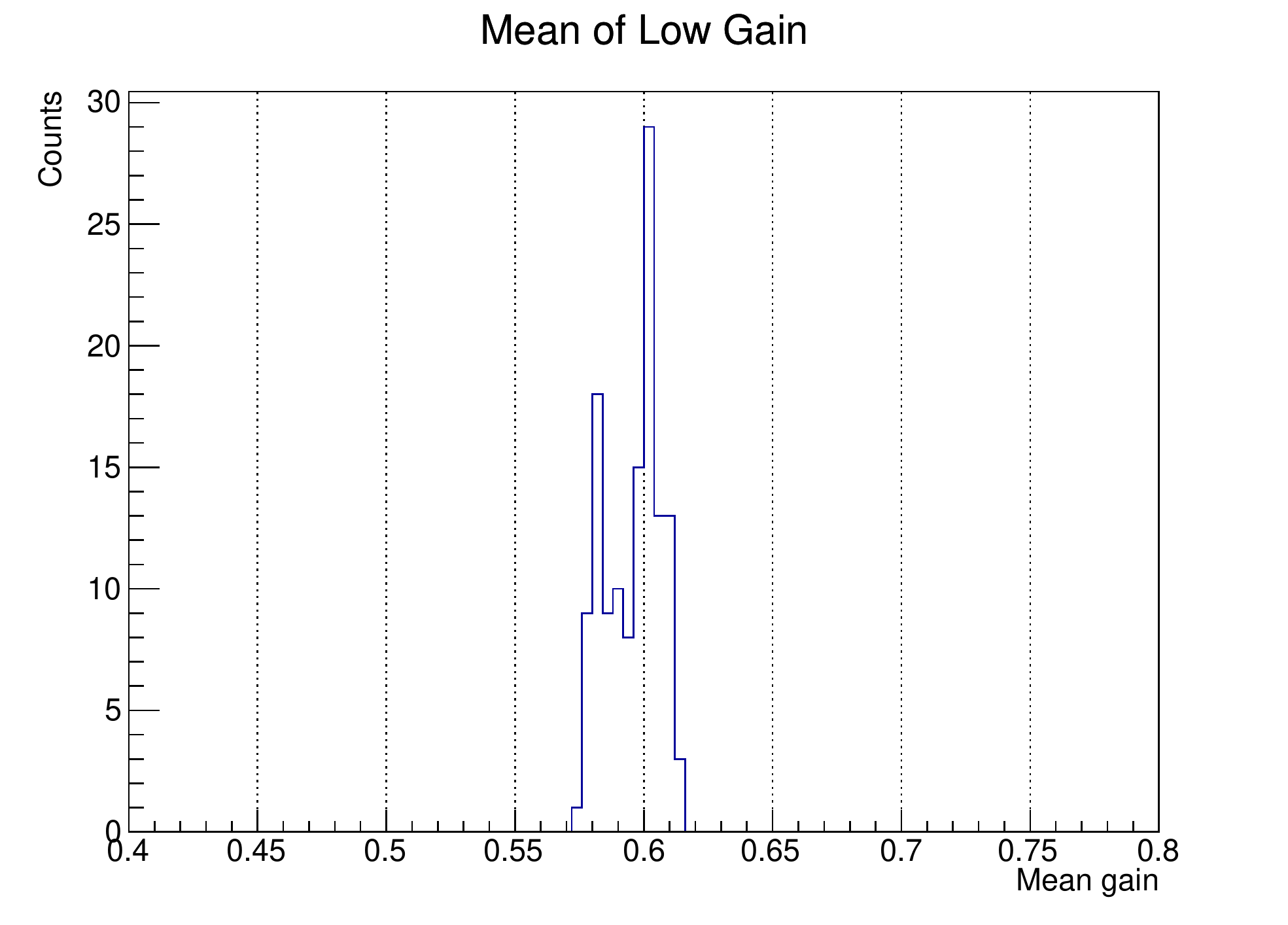}
\end{center}
\caption{\setlength{\baselineskip}{4mm} Distribution across channels and FEEs of the mean gain value for the high-gain (left) and low-gain (right) channels.}
\label{fig:FEE_gain}
\end{figure}

Measurements of the high and low gain signals were done using a PMT signal as test pulse. The PMT observed a bottle of liquid scintillator so that it would be stimulated by scintillation light produced by environmental background radiation. The PMT signal was split into two by a divider. One was directly recorded with an FADC, and the other was inserted into the FEE and both the high and low gain signals were recorded by the FADC with different channels. 
As shown in Fig.~\ref{fig:FEE_gain_fitting} a Gaussian was fit to the distribution of measured gain values to determine the mean gain for the high and low gain channels. By measuring the gain for all FEE channels, mean values were drawn in Fig.~\ref{fig:FEE_gain}. High gain is measured as ~16 and low gain is measured as ~0.6.

\subsection{Flash analog-to-digital converters}
JSNS$^{2}$ uses 28 CAEN v1721 500 MS/s FADC modules~\cite{cite:CAEN_FADC}. As shown in Fig.~\ref{fig:DAQ_scheme}, FADCs record PMT waveforms and it requires 216 channels which are equivalent to the 27 FADC modules. Additionally, we assigned one FADC as a trigger board which will be explained in the Sec.~\ref{sec:Trigger_board}. Four optical cables transfer data from the FADCs to the DAQ computer. The maximum data transfer rate allowed by the four optical cables is 320 MB/s. To permit 28 digitizers to be readout with four optical cables, 7 FADCs are linked with one optical daisy chain and connected to the DAQ PC by a CAEN A3818 PCIe card~\cite{cite:A3818}. Note that the 80 MB/s band width at maximum per optical cable comes from the CAEN Technical Information Manual~\cite{cite:CAEN_FADC_manual}. 

Each FADC is equipped with an internal clock, however the clocks across two separate FADC will, in general, be out of phase. We observed that even if separate FADCs were triggered from the same signal, they would readout with different trigger times. To remove this undesirable timing uncertainty, each FADC's clock is synchronized using a daisy chain method as described in Ref.~\cite{cite:FADC_timing_synchronization}.


\subsection{Trigger Board}
\label{sec:Trigger_board}
One FADC module is designated as a trigger board which generates global trigger to record the PMT waveforms in the all FADC modules. Whenever a NIM-type signal is inserted into any of the trigger board channel, it generates a NIM-type trigger from the TRG-out on the front panel. The trigger signal is fed into a digital fan-in/fan-out board which copies the signal to 32 outputs. The copied trigger signals are inserted into each FADC module including the trigger board itself. During the offline data production, a process on the KEKCC would inspects the recorded data from the trigger board and assigns a trigger-word to the event to indicate which channel produced the trigger. 

\subsection{Run control and monitoring}
\begin{figure}[h]
\begin{center}
\includegraphics[scale=0.6]{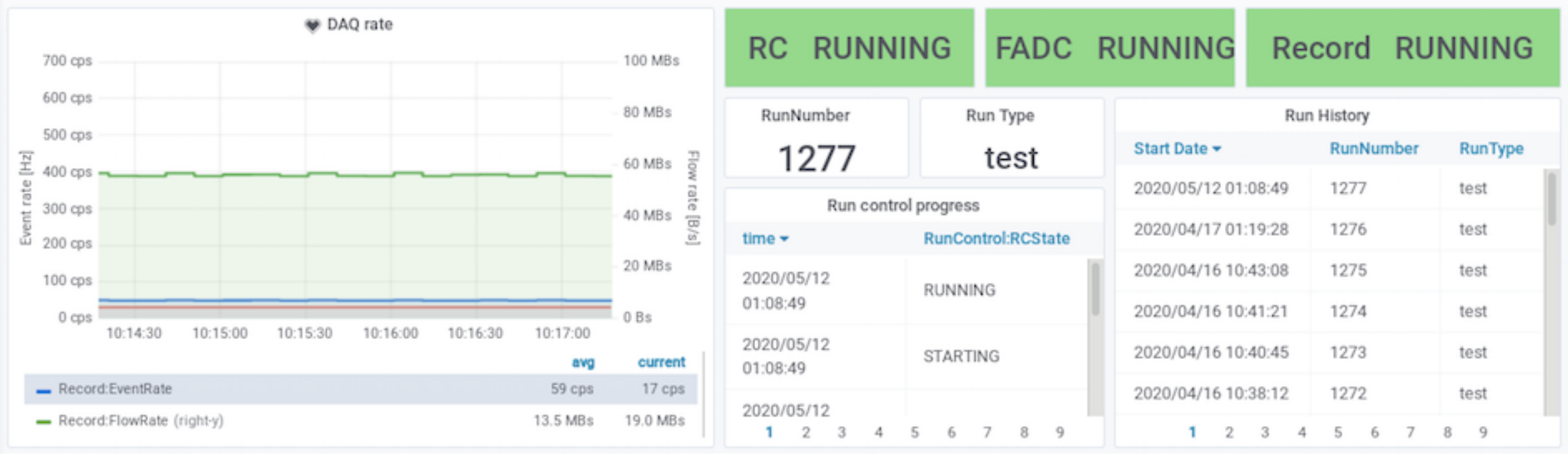}
\end{center}
\caption{\setlength{\baselineskip}{4mm}A screenshot of the DAQ monitoring system which is made by the Grafana. The left panel displays the trigger rate (green line) and transferred data size (blue line). The right panel displays current status of the DAQ including run-number and run-type.} 
\label{fig:run_monitoring}
\end{figure}

The run control program which is based on the Experimental Physics and Industrial Control System (EPICS)~\cite{cite:epics} controls the data taking. The run-number is loaded from a MySQL database automatically at the start of each new run. During each data taking run the run control program calculates values such as the transferred data size and number of triggers; the values calculated by the run control are periodically stored in a database. Figure~\ref{fig:run_monitoring} is a screenshot of the DAQ monitoring system which is produced by the Grafana~\cite{cite:Grafana}. The left panel shows the calculated trigger rate, and the right panel shows the status of the DAQ.

\subsection{Data acquisition}
As shown in Fig.~\ref{fig:DAQ_scheme}, the DAQ computer, Ubuntu operation system~\cite{cite:Ubuntu} with 8 CPUs of Inter Core i7-7700, 16 GByte memory, one ethernet connection, loads the data from the FADCs via four optical cables. The FADC data format is already decided in the module as described in the CAEN FADC V1721 documentation~\cite{cite:CAEN_FADC}. The DAQ PC adds the FADC serial number before each FADC data, header and tail magic number for each event, compresses each event, and transfers it to the DQM computer via a TCP/IP socket. An Adler-32 checksum method is used to confirm the data transfer to the DQM computer without fail.  

\subsection{Data quality monitoring}
\begin{figure}[h]
\begin{center}
\includegraphics[scale=0.7]{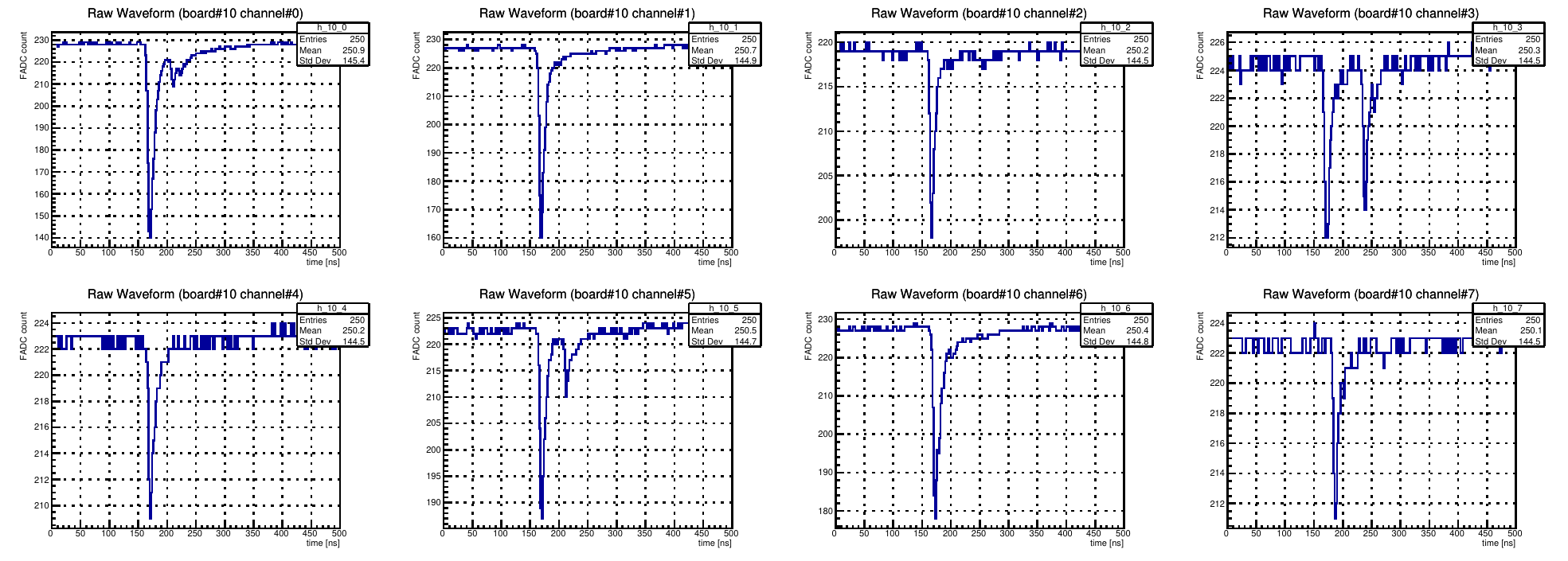}
\end{center}
\caption{\setlength{\baselineskip}{4mm}An example of the waveforms of 8 high-gain channels.} 
\label{fig:DQM_waveform}
\end{figure}

A program running on the DQM computer receives the data, decompresses each event, and performs quality checks. Low-level checks are done to ensure the data conforms transfer protocol, which includes two magic numbers embedded in the datastream and a checksum value. If there is no mismatch, the receiver program stores the data and makes each sub-run according to the data size.
Then, a sender program compresses and copies each sub-run of raw data to the KEKCC, where the data is converted to a ROOT file format for eventual analysis. At the same time, an analyzer process partially unpacks the raw data to update a waveform display. Figure~\ref{fig:DQM_waveform} shows a screenshot of the waveform display. 

\section{Trigger conditions}
\label{sec:trigger}
The main physics signal of the JSNS$^{2}$ comes from the collision of the 3 GeV proton beam accelerated by the J-PARC Rapid Cycling Synchrotron (RCS) with a mercury target~\cite{cite:JSNS2_proposal, cite:JSNS2_TDR}. Therefore, the JSNS$^{2}$ DAQ must be able to receive trigger signals from the beam facility so that signals from interesting physics events can be reliably detected. In order to know the beam timing, the scheduled timing from the J-PARC RCS accelerator is used. The scheduled timing is
produced by the Radio Frequency (RF) of the accelerator to inject and extract
proton beam of RCS~\cite{cite:Acc-timing}. One of scheduled timing to fire
the kicker magnets for the extraction of RCS beam to MLF and J-PARC Main Ring
(MR), which is coming to the power supplies of kicker magnets earlier than
the real beam timing by $\sim$100\,$\mu$s is also used by the JSNS$^2$.   
This scheduled timing is generated regardless that 3GeV beam is really coming 
to MLF, therefore we use signals from current transformers (CTs)
located to the beam line from the RCS accelerator to MLF~\cite{cite:3NBT} to
judge whether 3GeV beam is transported. Ten CTs are installed to measure the
beam intensity in the beam transporting line using the induced current of the
coil by the 3GeV proton beam~\cite{cite:CT}. The signals of CTs are
discriminted and transported to the experimental groups of MLF, and the
timing of the CT signal is $\sim$ 2.7\,$\mu$s after the real beam due to the
electronics delay, transportation of signal. JSNS$^{2}$ uses both the kicker trigger to take the beam triggered data and the CT trigger to confirm whether the beam was arrived or not. Besides the beam trigger, a self-trigger is required to take the delayed signal of the IBD as well as calibrate the detector with radioactive sources and Michel electron. 

\subsection{Beam-trigger Timing Check}
\begin{figure}[h]
\begin{tabular}{c c}

\begin{subfigure}{.4\linewidth}
\centering
\includegraphics[scale=0.5]{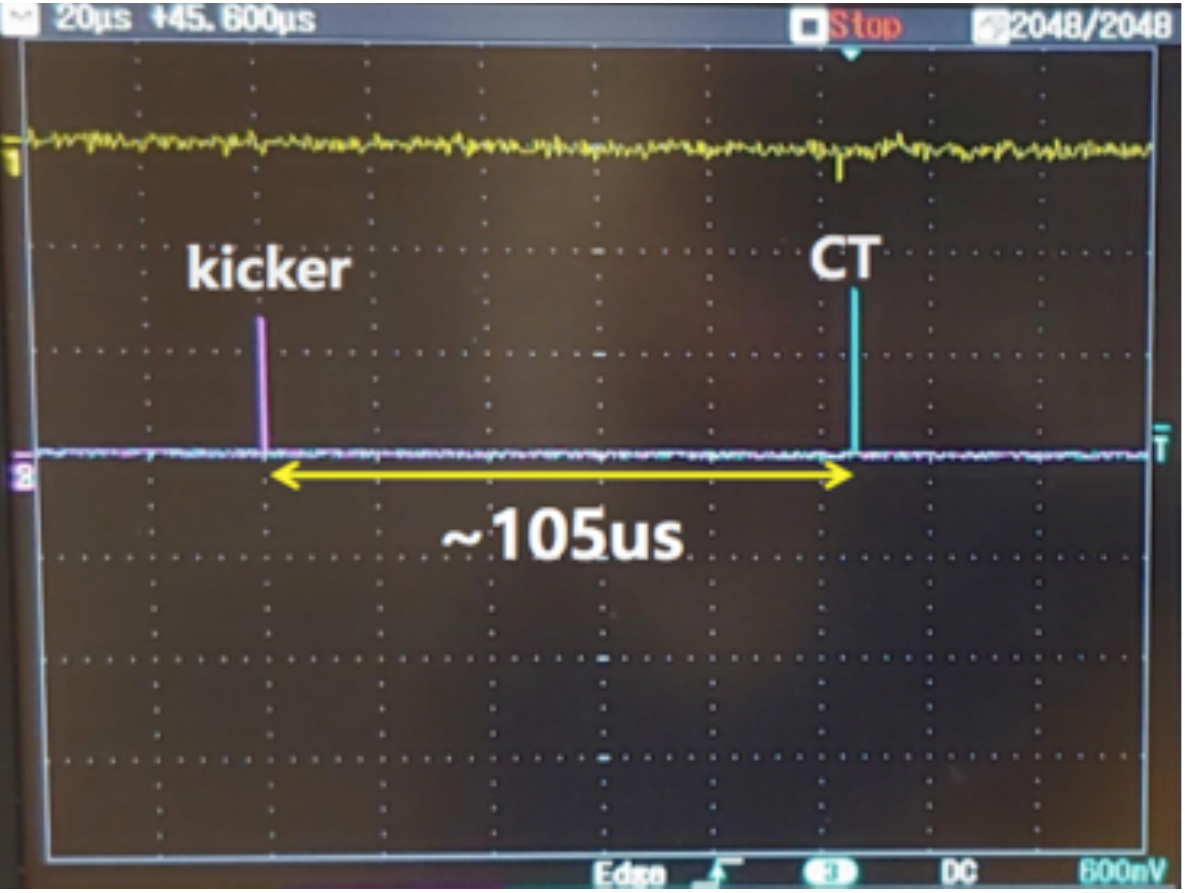}
\caption{Both the kicker and the CT triggers.}
\label{fig:beam_trigger_c}
\end{subfigure}
&
\begin{subfigure}{.6\linewidth}
\centering
\includegraphics[scale=0.5]{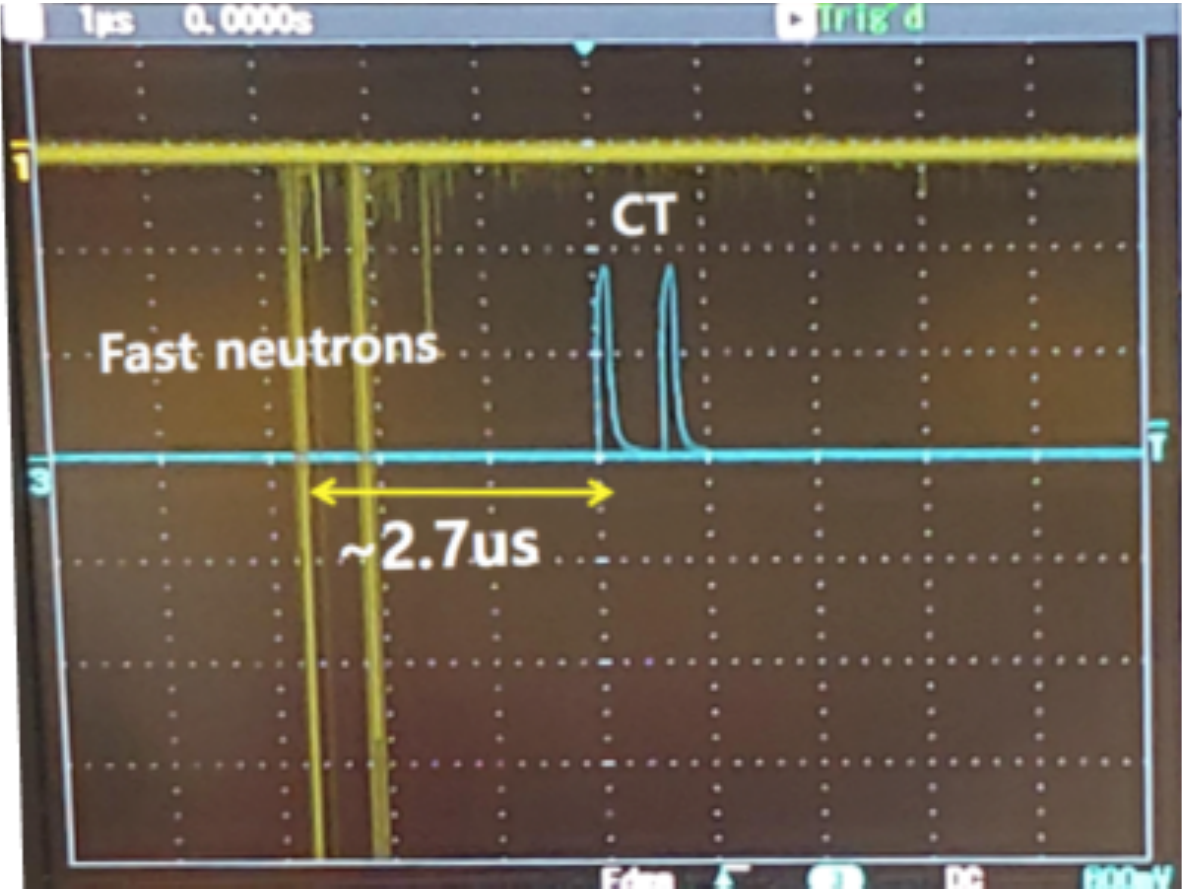}
\caption{Both the plastic scintillator signal and the CT trigger.}
\label{fig:beam_trigger_d}
\end{subfigure}
\end{tabular}
\caption{\setlength{\baselineskip}{4mm} (a) shows the timing separation between the kicker and the CT trigger which is around 105\,$\mu$s. (b) shows the timing separation between the plastic scintillator signal and the CT trigger which is around 2.7\,$\mu$s. The plastic scintillator signal comes from the fast neutrons which are produced when a proton collides the mercury target. We delayed the kicker trigger to adjust timing with the proton collision to take beam triggered data.} 
\label{fig:beam_trigger}
\end{figure}

\begin{figure}[h]
\begin{center}
\includegraphics[scale=0.5]{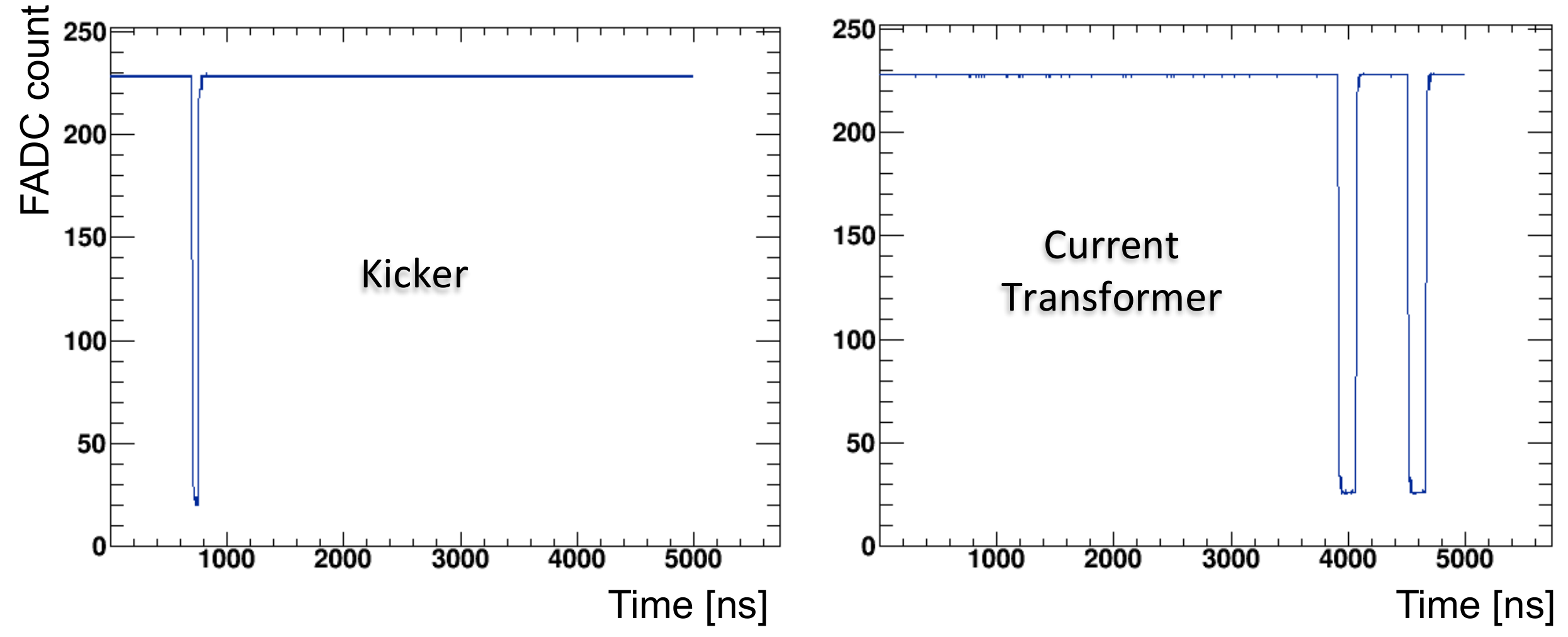}
\end{center}
\caption{\setlength{\baselineskip}{4mm}The stored waveform of the kicker (left) and the CT (right) which is taken with a kicker trigger. Note that we inserted the kicker and the CT into different channels and set the event window to 5$\mu$s.} 
\label{fig:beam_triggered_data}
\end{figure}

We performed measurements in order to confirm and evaluate the timing of the beam kicker and CT signals relative to observed fast neutron events. J-PARC provides two bunches of the beam, which are separated with around 600\,ns. We temporarily placed a plastic scintillator block with one PMT near upstream of the mercury target to check the real beam timing with respect to the kicker timing and the CT signal precisely. The shielding material in the direction of upstream is mainly made by concrete (not iron) and fast neutrons created by the beam halo can penetrate    
the shields~\cite{cite:NF2019}, therefore the plastic scintillator can observe beam induced fast neutrons. This was already measured by the background measurement of the JSNS$^2$ group performed in 2014~\cite{cite:PTEP}. Figure~\ref{fig:beam_trigger} displays each oscilloscope screenshot for both the kicker and the CT triggers(\ref{fig:beam_trigger_c}), and both the plastic scintillator signals and the CT triggers(\ref{fig:beam_trigger_d}). We delayed the kicker trigger to adjust timing with the proton collision to take beam triggered data. Figure~\ref{fig:beam_triggered_data} shows the stored waveform with the kicker trigger. Note that we inserted the kicker and the CT into different channels with a 5\,$\mu$s event window.

\subsection{Self-trigger}
\begin{figure}[h]
\begin{center}
\includegraphics[scale=0.5]{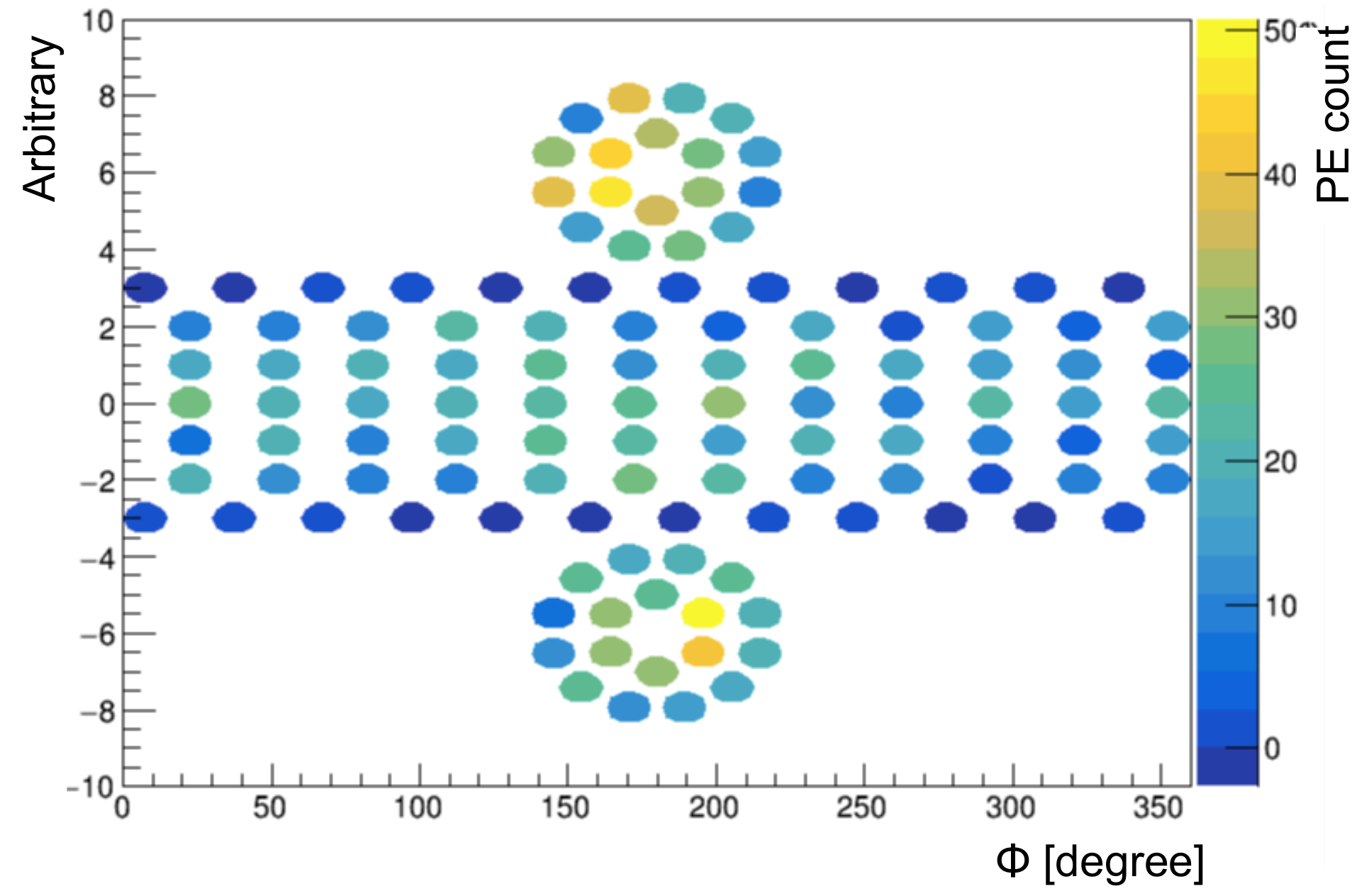}
\end{center}
\caption{\setlength{\baselineskip}{4mm}An event display of one event. Each circle denotes a PMT and the color shows the amount of the observed charge.}
\label{fig:LS_data}
\end{figure}

JSNS$^{2}$ plans to use an analog sum, as shown in Fig.~\ref{fig:DAQ_scheme}, rather than the number of PMT hits. The required energy threshold for most physics events of interest is several MeV, which will result in nearly all PMTs being hit by one or more photons. To test the effectiveness of the analog sum as our primary trigger condition we suspended a 100\,mL bottle of liquid scintillator in the center of the JSNS$^{2}$ detector. This test was done when the detector was otherwise empty. With around 2\,MeV threshold, we took self triggered data with an average event rate of 15\,Hz successively. Figure~\ref{fig:LS_data} shows an example event display of single event.

\section{Conclusion}
The JSNS$^{2}$ experiment uses CAEN FADCs to digitize and record PMT analog signals after the signals are copied and amplified with two different gains by FEEs. In addition to copying and amplifying the PMT signals the FEE also produces an analog sum which is used as part of the self-trigger system. Two PCs, a DAQ PC and a DQM PC, are used to separate the data taking and the data processing to allow for an overall higher data readout rate and to decrease over-burden. We have tested individual components of the DAQ and the system as a whole, demonstrating the DAQ's ability to operate using beam triggers as well as self triggers.

\section*{Acknowledgments}
We thank the J-PARC staff for their supports, especially MLF and accelerator staff. We acknowledge the support of the Ministry of Education, Culture, Sports, Science and Technology (MEXT) and the JSPS grants-in-aid (Grant Number 16H06344, 16H03967), Japan. This work is also supported by the National Research Foundation of Korea (NRF) Grant No. 2016R1A5A1004684, 2017K1A3A7A09015973,  2019R1A2C3004955, 2016R1D1A3B02010606, 2017R1A2B4011200 and 2018R1D1A1B07050425 funded by the Korea Ministry of Science and ICT. Our work have also been supported by a fund from the BK21 of the NRF. The University of Michigan gratefully acknowledges the support of the Heising-Simons Foundataion. This work conducted at Brookhaven National  Laboratory was supported by the U.S. Department of Energy under Contract DE-AC02-98CH10886. The work of the University of Sussex is supported by the Royal Society grant no.IES\textbackslash R3\textbackslash 170385.

\end{document}